\documentclass[10pt,letterpaper,twocolumn]{article}

\usepackage{ol2}
\usepackage[draft]{hyperref}
\usepackage{amsmath}

\begin{document}

\twocolumn[

\title{Optimal Concentration of Light in Turbid Materials}

\author{E.G. van Putten,$^{1,*}$ A. Lagendijk,$^{1,2}$ and A.P. Mosk$^1$}

\address{
$^1$Complex Photonic Systems, Faculty of Science and Technology and MESA$^+$ Institute for\\ Nanotechnology, University of Twente, P.O. Box 217, 7500 AE Enschede, The Netherlands \\
$^2$FOM Institute for Atomic and Molecular Physics, Science Park 104, 1098 XG Amsterdam, The
Netherlands\\
$^*$Corresponding author: E.G.vanPutten@utwente.nl
}

\begin{abstract}
In turbid materials it is impossible to concentrate light into a focus with conventional optics. Recently it has been shown that the intensity on a dyed probe inside a turbid material can be enhanced by spatially shaping the wave front of light before it enters a turbid medium. Here we show that this enhancement is due to concentration of light energy to a spot much smaller than a wavelength. We focus light on a dyed probe sphere that is hidden under an opaque layer. The light is optimally concentrated to a focus which does not exceed the smallest focal area physically possible by more than $68\%$. A comparison between the intensity enhancements of both the emission and excitation light supports the conclusion of optimal light concentration.
\end{abstract}
]

\noindent In turbid materials such as white paint, biological tissue, and paper, spatial fluctuations in refractive index cause light to be scattered. Scattering is seen as a huge vexation in classical imaging techniques where it degrades the resolving power.\cite{Sebbah1999aa} This decrease in resolution is caused by the fact that light carrying information about the fine spatial details of a structure has to travel further through the medium than the light carrying low spatial frequency information.\cite{Ishimaru1978} Due to the importance of imaging inside turbid materials, many researchers are trying to suppress turbidity.\cite{Huang1991aa,Schmitt1999aa,Denk1990aa,Helmchen2005aa,Leith1966aa,Yaqoob2008aa}

Although light scattering is detrimental to imaging, it is recently shown that scattering can be exploited to increase the amount of light energy deep inside turbid materials.\cite{Vellekoop2008aa} By spatially shaping the wave front of the incident light, the emission of a small dyed probe sphere hidden inside the turbid layer was strongly enhanced. Despite the fact that this enhancement proves an increase of excitation intensity at the probe position, it remains unclear what the spatial distribution of the excitation light is. From experiments with microwaves\cite{Simon2001aa} and ultrasound\cite{Fink1997aa} and recent far field experiments with light\cite{Vellekoop2009aa} it is known that scattering can be used to concentrate energy.

In this Letter we will experimentally show that we can also use scattering to focus light inside a turbid material to an optimal small spot, i.e, as small as it can physically be. The focus is created on a nano-sized dyed probe sphere hidden under a strongly scattering layer. A comparison between the intensity enhancements of the probe emission and the excitation light supports our conclusion of optimal light concentration.

\begin{figure}
  \centering
  \includegraphics[width=8.3 cm]{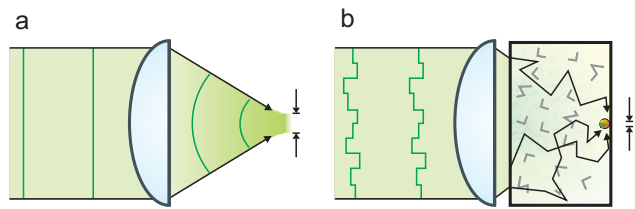}
  \caption{
  Principle of the experiment. (a) A positive lens focusses a plane wave to a spot which is limited in size by the numerical aperture (NA) of the lens. (b) A strongly turbid material behind the lens scatters the light so that no focus is formed. By matching the incident wave front to the scattering sample, we force constructive interference at a target position inside the sample. The light now arrives from all directions at the target position, significantly increasing the NA of the system.
  }
  \label{fig:SketchExperiment}
\end{figure}
Figure~\ref{fig:SketchExperiment} shows the principle of our experiment. (a) Ordinarily a positive lens focusses an incident plane wave to a spot with a size that is limited by the numerical aperture (NA) of the lens. (b) A strongly turbid material behind the lens scatters the light so that no focus is formed. By matching the incident wave front to the scattering sample, we force constructive interference at a target position inside the sample. At this position multiple scattered light arrives from all angles, significantly increasing the NA of the system. The focal size is no longer limited by the original lens, but can be minimized to the smallest spot size physically possible.

The possibility to focus light to a subwavelength spot inside scattering materials yields exciting opportunities. In biological imaging, for example, selective illumination of fluorescent areas with high resolution is highly desirable. The efficient light delivery to places inside scattering materials might also be used to study more fundamental properties of light transport in both ordered and disordered structures.

Our experiments are performed on opaque layers of strongly scattering zinc oxide (ZnO) pigment sprayed on top of a low concentration of dyed polystyrene spheres that will act as local intensity probes. We used probe spheres with a radius of R~=~$150$~nm and R~=~$80$~nm. ZnO is one of the most strongly scattering materials known and shows no fluorescence in the spectral region where the probes emit. The thicknesses of the scattering layers range between $7.5\pm1~\mu$m and $25\pm 4~\mu$m and have a mean free path of $\ell = 0.7 \pm 0.2~\mu$m. By measuring the angular resolved transmission through the ZnO layers\cite{Muskens2008aa}, we determined their effective refractive index $n_\text{eff}$ to be $1.35\pm0.15$.\cite{Vera1996,Rivas2003aa}

Using a wave front synthesizer, similar to the one discussed in Ref.~\cite{Vellekoop2008aa}, we spatially divide a monochromatic laser beam ($\lambda = 532$~nm) into up to $640$ square segments of which we individually control the phase. The shaped beam is focussed onto our sample using a microscope objective (NA~=~$0.95$). The same microscope objective is used to capture the fluorescence from a probe hidden under the scattering layer. At the back of the sample we use an oil-immersion microscope objective (NA~=~$1.49$) to directly image the excitation and emission light at the probe. A digital feedback system that monitors the amount of fluorescence, tailors the wave front to maximize the emission of a probe sphere hidden under the scattering layer. Both the fluorescence measured at the front and at the back of the sample were used independently to feed the digital feedback system and we did not observe a difference in our results. Because detecting fluorescence at the back of our sample allows us to use low shutter times on our camera, we used this signal as a feedback for most of our measurements.

\begin{figure}
  \centering
  \includegraphics[width=8.4 cm]{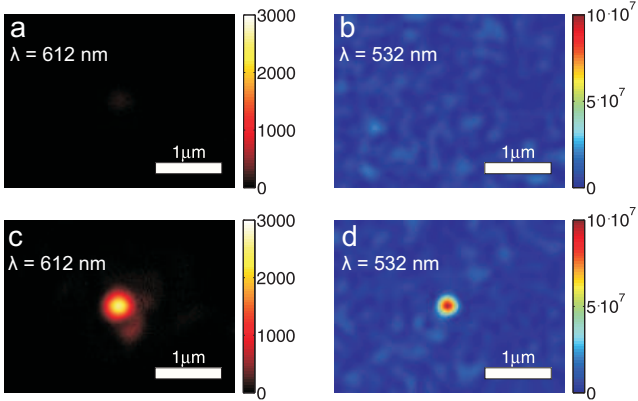}
  \caption{
  Fluorescence (a, c) and excitation (b, d) images taken at the position of a probe sphere hidden under a layer of zinc oxide. The probe sphere has a radius of R~=~$150$~nm. The images are taken at the back of the sample where the sphere is directly visible. In (a) and (b) we focus a plane wave onto the sample. Scattering creates a random specular pattern of excitation light resulting in a small amount of fluorescent response of the probe sphere. In (c) and (d) we illuminate the sample with a shaped wave which is created to maximize the fluorescent emission. Aside from the fluorescent emission enhancement we see a sharp focus of excitation light. All intensities are in counts/second.
  }
  \label{fig:Images}
\end{figure}
In Fig.~\ref{fig:Images} we see a typical result of the experiment. When we illuminate the sample with a focussed plane wave, we measure a low fluorescence response from a R~=~$150$~nm probe (a) and we see a speckle pattern in the excitation light (b). Nothing in the speckle pattern reveals the position of the probe. If a shaped wave, created to maximize the fluorescent emission, is focussed onto the sample we measure an intensity enhancement of the emission (c) and we see a sharp focus of excitation light at the position of the probe (d). It is surprising to see that this focus is smaller than the probe sphere.

\begin{figure}
  \centering
  \includegraphics[width=8 cm]{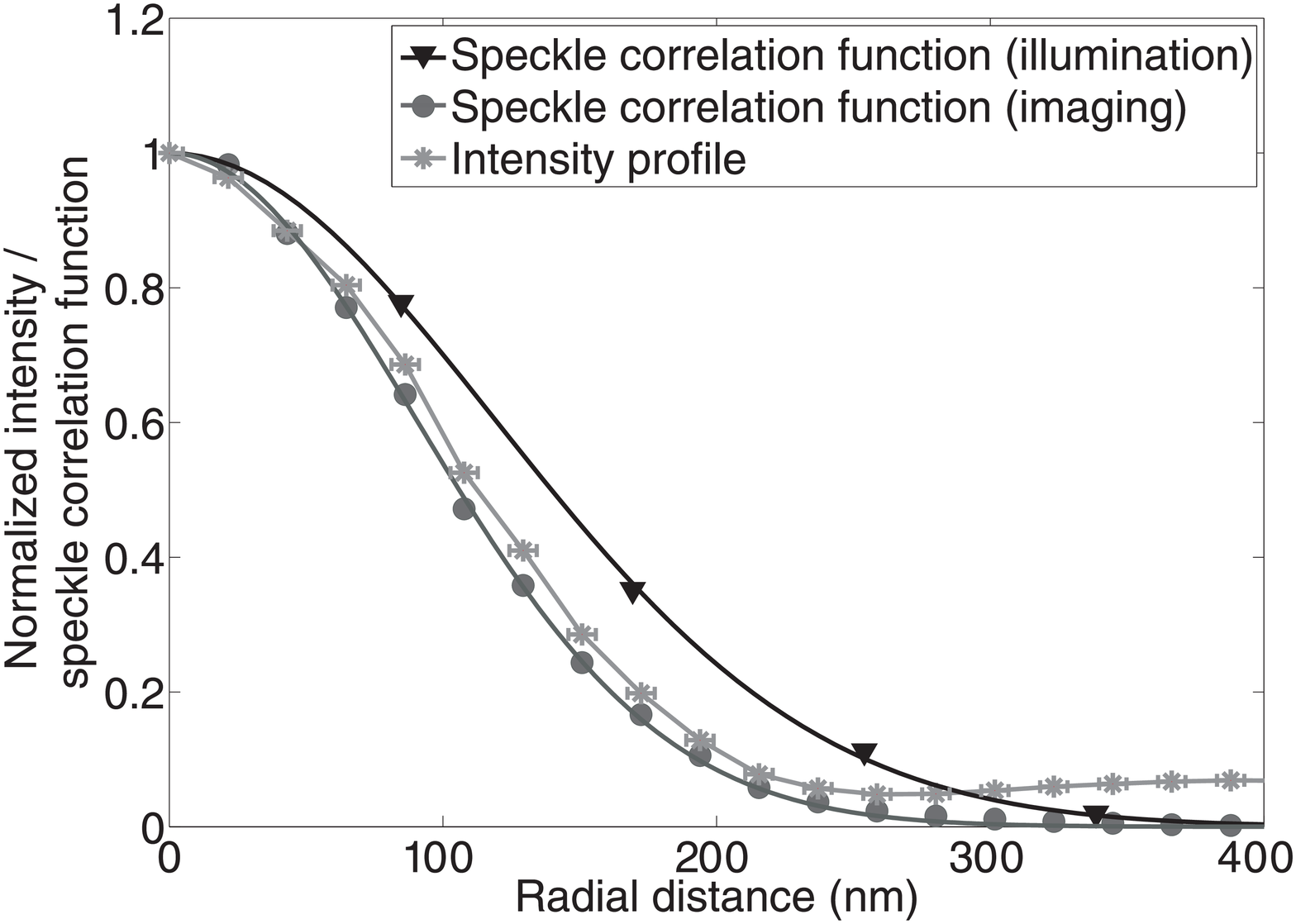}
  \caption{The measured radial intensity profile of the generated spot (stars) and the speckle correlation functions of the illumination (NA~=~$0.95$, triangles) and the imaging (NA~=~$1.49$, dots) microscope objectives.
  }\label{fig:Spotsize}
\end{figure}

The radial intensity profile in the focus is shown in Fig.~\ref{fig:Spotsize} together with the speckle correlation functions (SCFs) measured through both the illumination and the imaging microscope objective. The SCF is equal to the point spread function (PSF) of an optical system\cite{Cittert1934,Zernike1938,Goodman2000} giving the resolution limit of the illumination and imaging optics. The measured intensity profile has a half width at half max (HWHM) of $111\pm5$~nm and a peak intensity of $32.1$ times the average speckle background intensity. The size of the created spot is not limited by the illumination system, as it is substantially smaller than the measured illumination SCF. We also see that the measured size of the created spot is equal to the SCF of the imaging system, meaning that the created spot is smaller than or equal to the resolution limit of our microscope objective.

From literature we know that linear polarized light can be optimally concentrated to an elliptical focus.\cite{Richards1959aa} The HWHM of the two axis of this elliptical focus are given by $0.28\lambda / n_\text{eff}$ and $0.17\lambda / n_\text{eff}$ and the focal area by $0.048 \pi \lambda^2/n_\text{eff}^2$. In our experiment this would result in a smallest possible focal area of $0.02$~$\mu$m$^2$ with the HWHM of the two axes $110$~nm and $67$~nm. The resolution of our imaging microscope objective does not allow us to resolve the short axis of the spot. However, we see that without explicitly minimizing the spot size, the created spot area does not exceed the smallest spot area physically possible by more than $68\%$.

\begin{figure*}
  \centering
  \includegraphics[width=16 cm]{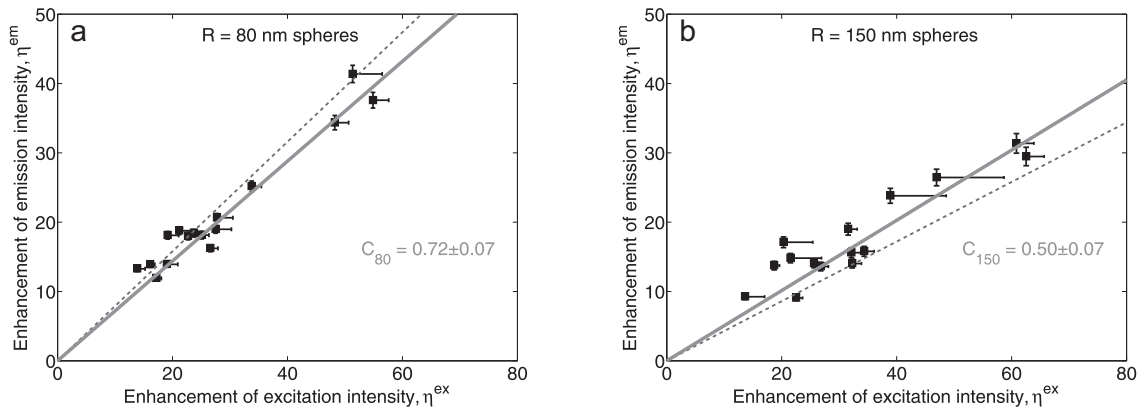}
  \caption{
  Measured enhancements of the excitation and emission intensity for spheres with a radius of (a) $R = 80$~nm and (b) $R = 150$~nm. The solid lines indicate the linear regression of the data points and the dashed lines represent the expected regression for light that is optimally concentrated to the smallest possible spot.
  }
  \label{fig:Enhancements}
\end{figure*}
To further investigate the created focus, we compare the intensity enhancements of the emission $\eta^\text{em}$ and excitation $\eta^\text{ex}$. For the excitation intensity, the enhancement is defined as the ratio between the peak intensity of the focus and the average diffusive background intensity. The diffusive background intensity is determined by averaging the intensity at the probe sphere over $100$ random realizations of the incoming wave front. The emission intensity enhancement is defined as the total emission when a focus is created divided by the average emission during the reference measurement. In Fig.~\ref{fig:Enhancements} we have plotted the measured enhancements for (a) $R = 80$~nm and (b) $R = 150$~nm probe spheres. For the same probe size the emission enhancements $\eta^\text{em}$ are proportional to $\eta^\text{ex}$. The proportionality constant is clearly different for the two probe sizes. The number of control segments was varied to create a large spread in enhancements.

We now theoretically investigate the relation between $\eta^\text{em}$ and $\eta^\text{ex}$. The emission power caused by a focused excitation field is proportional to the focus intensity integrated over the probe sphere volume. The reference emission power scales with the volume of the probe sphere. On the contrary, the excitation enhancement is independent of the probe volume and is determined by dividing the peak intensity of the focus by the reference speckle intensity. The ratio of the enhancements for a probe sphere with radius $R$, $C_R$, is now given by
\begin{equation}
    C_R \equiv
    \frac{\eta^\text{em}}{\eta^\text{ex}}
    = \frac{1}{V}\int_R{\frac{I}{I_\text{peak}} dV},
    \label{eq:Enhancements}
\end{equation}
where $V$ is the volume of the probe sphere.

Assuming that plane polarized light is optimally focussed to the smallest spot area\cite{Richards1959aa} in the center of the probe, we numerically calculate the overlap integral in Eq.~\ref{eq:Enhancements}. From the calculations we find that $C_{80} = 0.77$ for the $80$~nm spheres and $C_{150} = 0.43$ for the $150$~nm spheres (dashed lines in Fig.~\ref{fig:Enhancements}). These values are in good agreement with the experimental values $C_{80}$ and $C_{150}$ that we find from the linear regression of the data points, $C_{80} = 0.72\pm0.07$ and $C_{150} = 0.50\pm0.07$ (solid lines in Fig.~\ref{fig:Enhancements}). This data therefore support the conclusion that the light is being optimally concentrated.

In conclusion, we have focused light onto fluorescent probes hidden by a strongly scattering layer of zinc oxide by spatially shaping the phase of a light beam. We studied the shape and dimensions of the created focus. We found that the focus is optimally concentrated to a focus whose area is for certain within $68\%$ of the smallest focal area physically possible. A study of the intensity enhancements of both the fluorescence and excitation, performed on different probe sizes, supports the conclusion of optimal light concentration.

The authors thank Ivo Vellekoop and Willem Vos for support and valuable discussions. We thank Timmo van der Beek for help with the angular resolved transmission measurements. A. P. Mosk is supported by a VIDI grant from NWO.

%------------------------------------------------------------------------------------------------------------------------


\begin{thebibliography}{99}
    \bibitem{Sebbah1999aa}
    P. Sebbah, ``\textit{Waves and Imaging through Complex Media},'' (Kluwer Academic Publishers, 1999)

    \bibitem{Ishimaru1978}
    A. Ishimaru, ``Limitation on image resolution imposed by a random medium,'' Applied Optics {\bf 17,} 348--352 (1978)

    \bibitem{Huang1991aa}
    D. Huang, E.A. Swanson, C.P. Lin, J.S. Schuman, W.G. Stinson, W. Chang, M.R. Hee, T. Flotte, K. Gregory, C.A. Puliafito, and J.G. Fujimoto, ``Optical coherence tomography,'' Science {\bf 254,} 1178-1180 (1991)

    \bibitem{Schmitt1999aa}
    J.M. Schmitt, ``Optical coherence tomography (OCT): a review,'' IEEE Journal of Selected Topics in Quantum Electronics {\bf 5,} 1205-1215 (1999)

    \bibitem{Denk1990aa}
    W. Denk, J.H. Strickler, and W.W. Webb, ``Two-photon laser scanning fluorescence microscopy,'' Science {\bf 248,} 73-76 (1990)

    \bibitem{Helmchen2005aa}
    F. Helmchen and W. Denk, ``Deep tissue two-photon microscopy,'' Nature Methods {\bf 2,} 932-940 (2005)

    \bibitem{Leith1966aa}
    E.N. Leith, and J. Upatnieks, ``Holographic Imagery Through Diffusing Media,'' J. Opt. Soc. Am. {\bf 56,} 523-523 (1966)

    \bibitem{Yaqoob2008aa}
    Z. Yaqoob, D. Psaltis, M.S. Feld, and C. Yang, ``Optical phase conjugation for turbidity suppression in biological
	samples,'' Nature Photonics {\bf 2,} 110-115 (2008)

    \bibitem{Vellekoop2008aa}
    I.M. Vellekoop, E.G. van Putten, A. Lagendijk, and A.P. Mosk, ``Demixing light paths inside disordered metamaterials,'' Opt. Expr. {\bf 16,} 67--80 (2008).

    \bibitem{Simon2001aa}
    S.H. Simon, A.L. Moustakas, M. Stoytchev, and H. Safar, ``Communication in a Disordered World,'' Physics Today {\bf 54,} 38--43 (2001)

    \bibitem{Fink1997aa}
    M. Fink, ``Time-reversed acoustics,'' Phys. Today {\bf 50} 34--40 (1997)

    \bibitem{Vellekoop2009aa}
    I.M. Vellekoop, A. Lagendijk, and A.P. Mosk, ``Exploiting disorder for perfect focusing,'' Nature Photon. (2010), doi:10.1038/nphoton.2010.3


    \bibitem{Muskens2008aa}
    O.L. Muskens and A. Lagendijk, ``Broadband enhanced backscattering spectroscopy of strongly scattering media,'' Opt. Express {\bf 16} 1222--1231 (2008)

    \bibitem{Vera1996}
    M.U. Vera and D.J. Durian, ``Angular distribution of diffusely transmitted light,'' Phys. Rev. E {\bf 53} 3215--3224 (1996)

    \bibitem{Rivas2003aa}
    J.G. Rivas, D.H. Dau, A. Imhof, R. Sprik, B.P.J. Bret, P.M. Johnson, T.W. Hijmans, and A. Lagendijk, ``Experimental determination of the effective refractive index in strongly scattering media,'' Optics Commun. {\bf 220} 17--21 (2003)

    \bibitem{Cittert1934}
    P. H. van Cittert, ``Die wahrscheinliche Schwingungsverteilung in einer von einer Lichtquelle direkt oder mittels einer Linse beleuchteten Ebene,'' Physica {\bf 1,} (1934), as discussed in Ref.~\cite{Goodman2000}.

    \bibitem{Zernike1938}
    F. Zernike, ``The concept of degree of coherence and its application to optical problems,'' Physica {\bf 5,} 785 (1938), as discussed in Ref.~\cite{Goodman2000}.

    \bibitem{Goodman2000}
    J.W. Goodman, ``Statistical optics,'' Wiley, New York, (2000).

    \bibitem{Richards1959aa}
    B. Richards and E. Wolf, ``Electromagnetic Diffraction in Optical Systems. II. Structure of the Image Field in an Aplanatic System,'' Proc. R. Soc. London, Ser. A {\bf 253} 358--379 (1959)
    \end{thebibliography}
\end{document}